# Enhanced Doppler Effect in the Upper Hybrid Resonance Microwave Backscattering Experiment

A.B. Altukhov, A.D. Gurchenko, E.Z. Gusakov, L.A. Esipov, S.I. Lashkul, A.Yu. Stepanov

*Ioffe Institute, Politekhnicheskaya 26, 194021 St.Petersburg, Russia*

Observations of enhanced Doppler frequency shift effect of the highly localized microwave backscattering in the upper hybrid resonance are reported. The experiment is performed at FT-2 tokamak, where a steerable focusing antenna set, allowing off equatorial plane plasma extraordinary wave probing from high magnetic field side, was installed. A separate line less than 1.5 MHz wide and shifted by up to 2 MHz is routinely observed in the backscattering spectrum under condition of accessible upper hybrid resonance. The enhanced frequency shift is explained by the growth of poloidal wave number of the probing wave in the resonance. The new scheme for local diagnostics of fluctuations poloidal rotation based on this effect is proposed.

## 1. INTRODUCTION

Investigation of tokamak plasma poloidal rotation attracts considerable attention nowadays because of the stabilizing influence of its inhomogeneity on several types of gradient instability and suppression of associated anomalous transport [1]. The Doppler frequency shift of backscattering (BS) signal at oblique microwave plasma probing in the presence of cut off (Doppler reflectometry) is often used for diagnosing of poloidal plasma velocity in magnetic fusion devices [2 – 4]. The typical value of frequency shift of BS microwave of several hundred kHz in these diagnostics observed in ohmic discharge is usually substantially smaller than its broadening, which complicates interpretation and reduces the accuracy and time resolution of measurements.

In the present paper a possibility of a drastic increase of the Doppler frequency shift of microwave BS signal in toroidal devices, based on the upper hybrid resonance (UHR) BS is demonstrated experimentally. The UHR BS or enhanced scattering [5] utilizes for local diagnostics of small-scale plasma fluctuations the growth of wave vector and electric field of the probing extraordinary (X-mode) wave in the UHR, where condition $f_i^2 = f_{ce}^2(R) + f_{pe}^2(r)$ is fulfilled for the probing frequency $f_i$ ($R$ and $r$ are tokamak major and minor radii, $f_{ce}$ and $f_{pe}$ are electron cyclotron and plasma frequencies, correspondingly). To provide the UHR accessibility in tokamak experiment the probing wave is launched from the high magnetic field side of the torus under conditions when the electron cyclotron resonance layer exist somewhere in plasma. This technique is only sensitive to fluctuations possessing wavelength smaller than half-probing wavelength. The scattering cross section of the UHR BS $\rho$, shown in Fig. 1 for FT-2 tokamak parameters, experiences very sharp maximum at the fluctuation wave number $q_{conv} \equiv 2(2\pi f_i/c)\sqrt{c/V_{Te}}$, which corresponds

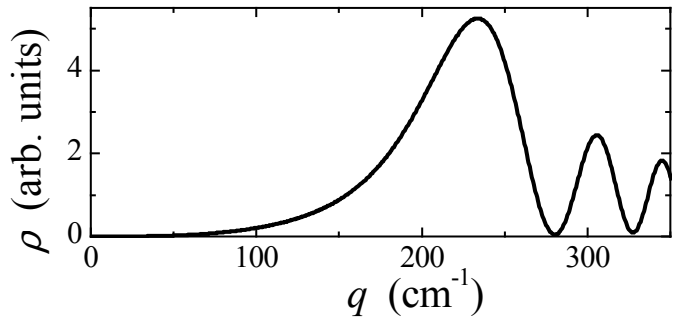

*Fig. 1.* – *BS efficiency versus fluctuation wave number perpendicular to UHR.*

to BS of the probing wave in its linear conversion point [5]. According to [6, 7], in toroidal devices, where the UHR and magnetic surfaces do not coincide due to dependence of magnetic field on the major radius $R$ (see Fig. 2), the large probing wave vector, perpendicular to the UHR surface, has a finite projection onto the poloidal direction. In the case of BS cross-section maximum this projection is given by relation

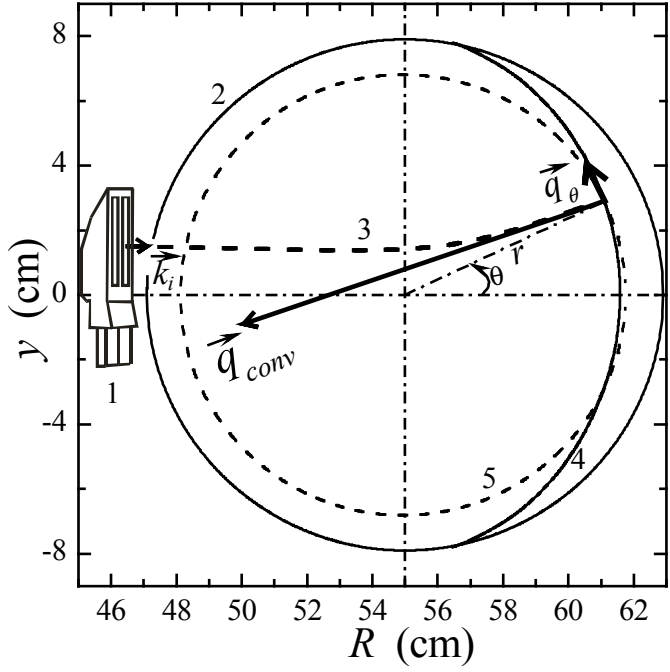

$$k_\theta = k_{\theta 0} - \frac{q_{conv}}{2} \frac{\vec{e}_\theta \vec{e}_R f_{ce}^2}{R|\vec{\nabla}(f_{pe}^2 + f_{ce}^2)|}\bigg|_{UHR}, \quad (1)$$

where $k_{\theta 0}$ gives the probing X-mode poloidal wave number out of the UHR zone, $\vec{e}_\theta$ and $\vec{e}_R$ are unit vectors in poloidal and major radius directions. In the outer discharge region close to the circular limiter, we can transform this expression to the following simplified form [6, 7]

*Fig. 2.* – FT-2 tokamak poloidal cross section. 1 – Antennae, 2 – limiter, 3 – central ray of the probing beam, 4 – UHR surface, 5 – magnetic surface, $\vec{k}_i$ – incident wave vector in vacuum, $\vec{q}_{conv}$ – fluctuation wave vector at maximum of BS efficiency, $\vec{q}_\theta$ – poloidal projection of $\vec{q}_{conv}$.

$$k_\theta \approx k_{\theta 0} + \frac{q_{conv}}{2} \frac{y}{r} \frac{f_{ce}^2}{f_{pe}^2} \frac{L_n}{R}\bigg|_{UHR}, \quad (2)$$

where $y$ and $r$ are a vertical displacement from the equatorial plane and minor radius of BS point at the UHR surface, correspondingly, and $L_n$ is the density scale length. This projection, which can be much larger than the poloidal component of wave vector at the antenna, can lead to substantial enhancement of the Doppler frequency shift of the microwave BS by fluctuations moving with poloidal plasma flow. The frequency shift corresponding to the BS cross section maximum, according to (1), is given by

$$f_D = 2\left[k_{\theta 0} + \frac{q_{conv}}{2} \frac{\vec{e}_\theta \vec{e}_R f_{ce}^2}{R|\vec{\nabla}(f_{pe}^2 + f_{ce}^2)|}\bigg|_{UHR}\right] V_\theta, \quad (3)$$

where $V_\theta = V_{ph} + V_{E \times B}$ is the fluctuation poloidal velocity in the laboratory reference system, $V_{E \times B}$ is the plasma drift velocity and $V_{ph}$ is the fluctuation phase velocity in the reference system moving with plasma. The BS spectrum width $\delta f$ in this model is determined by the diameter of the illuminated spot $\delta y$ as well as by the width of the scattering efficiency dependence on fluctuation wave number $\delta q$, shown in Fig.1. It is estimated by relation

$$\delta f \approx \left( \frac{\delta y}{\sqrt{2} y} + \frac{\delta q}{q_{conv}} \right) f_D \qquad (4)$$

## 2. EXPERIMENTAL RESULTS AND DISCUSSION

In this paper, the first observations of the giant Doppler frequency shift effect of the highly localized microwave BS in the UHR are reported. The experiment was performed at small research FT-2 tokamak possessing major radius $R$ = 55 cm and limiter's radius $a$ = 7.9 cm. The measurements were carried out at toroidal field $B_t \approx 2.2$ T, plasma current $I_p \approx$ 32 kA and density in range: $3.5 \times 10^{19}$ m$^{-3}$ < $n_e$ < $4.5 \times 10^{19}$ m$^{-3}$. Typical discharge duration was 90 ms. The microwave BS experiment was performed with a new steerable focusing double antennae set, shown schematically in Fig. 2, allowing off equatorial plane plasma X-mode probing from high magnetic field side. The maximal vertical displacement of antennae center is $y_a$ = ±2 cm, whereas the diameter of the wave beam at the position of UHR, as measured in vacuum is 1.5 – 1.7 cm, depending on the probing frequency. The coupling of emitting and receiving antennae is less than -35 dB. The probing was performed at power level of 20 mW in the frequency range 53 – 69 GHz providing UHR position scanning from $R \approx$ 60 cm to 63 cm. The BS signal was analyzed with quadrature scheme, which allows measurements of BS signal phase and amplitude, as well as determination of BS power spectra. Just after the new antennae set installation, a separate line shifted by up to 2 MHz became reliably observable in the BS spectrum under condition of accessible UHR. The BS power spectrum possessing the highest (2 MHz) shift, observed at +2 cm antenna vertical displacement, is shown in Fig. 3(a) ($f_s$ – scattering frequency). The amplitude of BS line there is higher than for the probing line because of the enhancement of scattering cross section and small coupling of antennae horns. The ratio of the line frequency shift and broadening is comparable to unity, which allows reliable determination of shift with high accuracy. The BS spectrum observed in equatorial plane and possessing no shift is shown in Fig. 3(b). As it is seen the width of this spectrum is smaller than in the case of Fig. 3(a), which is in agreement with the estimation (4). For comparison the BS spectrum observed with the same antenna set, shifted by +2 cm from equatorial plane, under conditions when the UHR is not accessible is shown in Fig. 3(c). This spectrum, which, in fact, corresponds to Doppler reflectometry with tilting angle of 14°, is only slightly shifted. Its frequency shift can be estimated with poor accuracy at the level of 100 kHz.

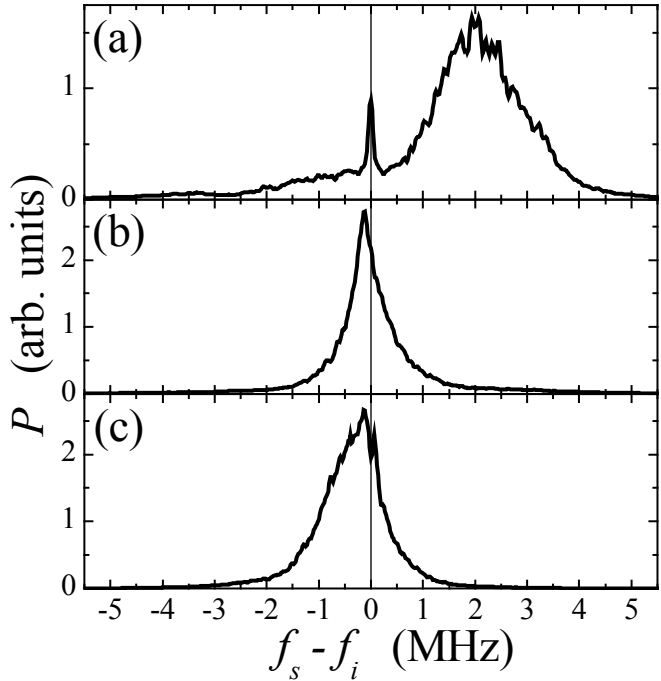

*Fig. 3.* – BS power spectra in (a) UHR ($y_a = +2$ cm), (b) UHR (-0.2 cm), (c) cut off (+2 cm).

The linear frequency shift proportionality to the displacement of BS point from the equatorial plane is confirmed in a special experiment in which the antennae vertical position was varied from discharge to discharge (see Fig. 4) at different radial UHR positions ($R = 60.1$ cm – circles; 61.7 cm – triangles). As it is seen, the frequency shift of the BS satellite changes sign, when the antennae set crosses the equatorial plane. The spectrum width is minimal at the equatorial plane probing and increases with absolute value of antennae vertical displacement in reasonable agreement with the estimation (4), predicting linear growth.

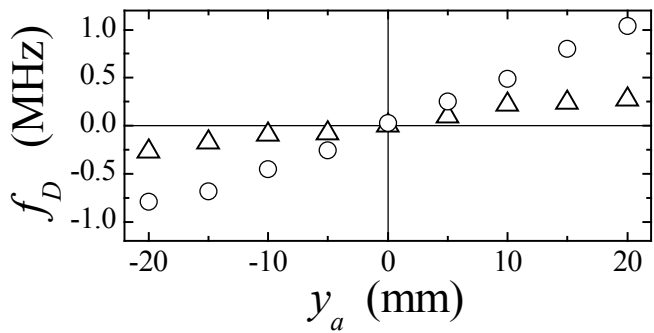

*Fig. 4.* – Doppler frequency shift versus antennae vertical displacement.

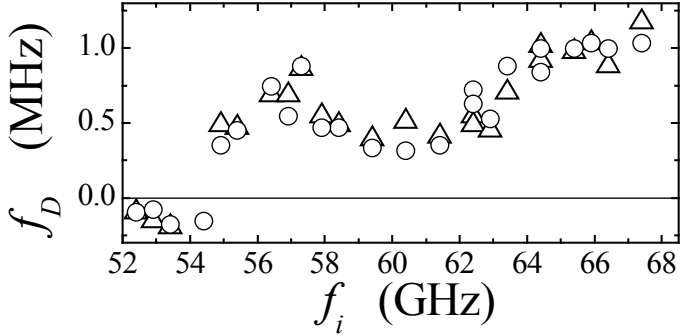

*Fig. 5. – Doppler frequency shift versus incident frequency ($y_a = +15$ mm). Circles – 30 ms, triangles – 33 ms.*

As it is shown in Fig. 5 the UHR BS satellite frequency shift appears to be strongly dependent on the probing frequency. It is positive and large at higher probing frequencies, corresponding to the inner position of the UHR and changes sign at the lower frequency corresponding to the edge location of the UHR. Dependences of fluctuation poloidal velocity on major radius in the equatorial plane, determined from Fig. 5 using formula (3) and the UHR condition, are shown in Fig. 6 for different discharge time, at which magnetic field, plasma current, density profiles, were slightly different. As it is seen, the poloidal velocity is positive in the inner plasma region, which is probably associated with plasma drift in negative radial electric field. This sign, as well as the velocity value of 2.5 km/s is in agreement with predictions of neoclassical theory [8] formula

$$E_r \approx \frac{T_i}{e}\left[\frac{\partial(\ln n_e)}{\partial r} + (1-k)\frac{\partial(\ln T_i)}{\partial r}\right],$$ (coefficient $k$ here depends on plasma collisionality

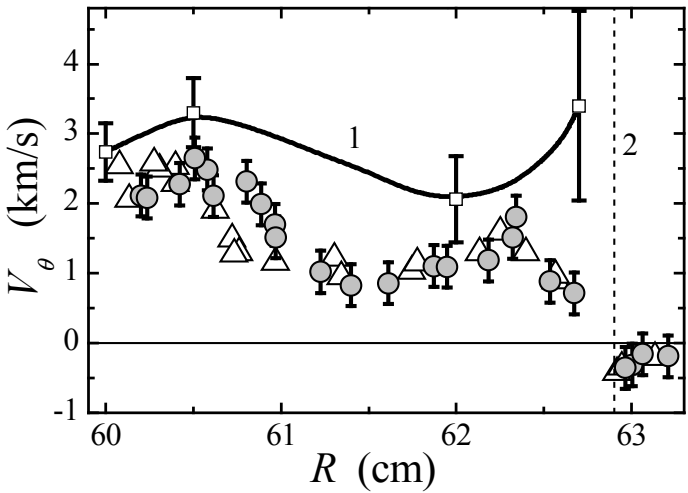

*Fig. 6. – Poloidal velocity profiles. Circles – 30 ms, triangles – 33 ms, 1 – neo-classical dependence, 2 – limiter position.*

($k(R \approx 61.5 \, \text{cm})$ = -0.50)). This expression is obtained neglecting the toroidal rotation contribution, which should not be large in FT-2 ohmic discharge at the safty factor *q(a)*=5. The theoreticaly predicted value of radial electric field in this region is negative, which corresponds to the dominance of ion channel of particle losses. The dependence of neoclassical poloidal rotation velocity on major radius, determined using experimental ion temperature and density profiles is also shown in Fig. 6. As it is seen, in the inner plasma region the neoclassical curve is close to the experimental values. The poloidal velocity decreases by a factor 2.7 at $R \approx 61.5$ cm, but at the plasma edge the experimental value of the poloidal velocity increases again towards last closed field surface (LCFS) in agreement with theoretical expectations. In the nearest vicinity of LCFS, at $R \approx 62.4$ cm the rotation velocity quickly decreases and changes sign in the edge region, where it is natural to expect positive plasma electric field, caused by fast electron losses along open magnetic field lines.

It is important to note that the calculated values of electron diamagnetic drift velocity all over the measurements region exceed the experimental values of fluctuations poloidal rotation velocity by a factor of 3 – 5. It means that the contribution of the fluctuation phase velocity $V_{ph}$ to the rotation, if any, is much smaller than one could expect for the electron drift eigen mode, possessing the spatial scale much smaller than ion Larmor radius. Based on this result we can conclude that the major part of the observed UHR BS signal is produced by non resonant fluctuations, probably resulted from nonlinear evolution of ion drift turbulence and its cascading to small scales. This conclusion is quite natural also because the fluctuations producing BS in the UHR should possess radial wave number much higher than the poloidal one and thus are not similar to the electron drift wave eigen-modes [1].

### 3. CONCLUSIONS

The giant Doppler frequency shift is observed in the off equatorial plane UHR BS experiment at FT-2 tokamak in agreement with theoretical predictions. The enhancement of the BS signal frequency shift is explained by growth of poloidal wave number of probing wave in the UHR, which should occur in toroidal devises, where the UHR and magnetic surfaces do not coincide and thus the huge UHR wave vector, which is perpendicular to the UHR surface, have a big projection onto poloidal direction. The fluctuation's poloidal velocity estimated from the frequency shift is close to the neoclassical theory expectations and much smaller than electron diamagnetic drift velosity. The last result favors the conclusion that the observed frequency shift is rather associated with the plasma $V_{E\times B}$ drift flow than with fluctuation phase velocity, which should be checked by spectroscopic and Doppler reflectometry measurements in preparation now.

The found robust effect has a potential for development of a new scheme for precise diagnostics of poloidal rotation of plasma fluctuations in tokamaks and stellarators possessing high spatial and temporal resolution.

ACKNOWLEDGMENTS

Supports of INTAS grants YSF2002-104, INTAS 01-2056, NWO-RFBR grant 047.016.015, RFBR grants 04-02-16534, HIII-2159.2003.2 are acknowledged.